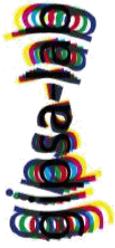
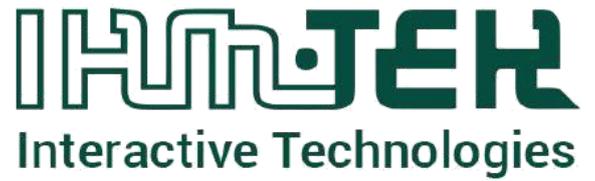

GIPSA-lab (Université Grenoble Alpes, CNRS, Grenoble INP), IHMTEK

# Analysis of tagging latency when comparing event-related potentials

Gregoire Cattan, Anton Andreev, Bastien Maureille, Marco Congedo
06/12/2018

# I. Introduction

Event-related potentials (ERPs) are very small voltage potentials produced by the brain in response to external stimulation and are measured with an electroencephalogram (EEG). Differences in the onset time and amplitude of ERPs reflect different sensory and high-level brain processing functions, such as the recognition of symbols (e.g., words or figures) or of the correctness of presented information, or changes in a subject's attention (1). For these reasons, ERPs are a useful tool for describing the processing of information inside the brain, with practical applications, for example, in the domain of brain-computer interfaces (BCIs), which allow for direct communication between the brain and an electronic device, bypassing the usual muscular and peripheral nerve pathways (2). The first BCI was designed by Vidal in the late 1970s (3) and allowed control of a cursor on a computer screen using an EEG signal. Recent improvements in EEG signal processing (4,5) have made the detection of ERPs more reliable, opening a promising field of BCI applications for people with disabilities (6–8) as well as for the general public (e.g., 9,10). In order to detect and evaluate an ERP in an ongoing electroencephalogram (EEG), it is necessary to tag the EEG with the exact onset time of the stimulus. A fixed latency engenders a constant offset which can be easily removed. Variance in latency, referred to as *jitter*, is more problematic, since it invalidates common operations such as trial averaging. Failing to control sequencing in the tagging pipeline causes problems when interpreting latency and can lead to contradictory conclusions (11–13). In this work, we present number of technical aspects which can influence latency.

The remainder of this report is organised as it follows: Section II describes a model for estimating latency. It contains A) the description of the tagging pipelines, B) an estimation of the latency for each tagging pipeline, and C) these estimations applied to the particular case of stimuli displayed in a matrix. Section III contains specific considerations on the case of a stimulus rendered by multiple cameras. Section IV reports on an example experiment. Section V presents our conclusions and recommendations.

# II. Model for estimating latency

## A. Tagging pipelines

On most platforms, we separate software rendering (*SoR*) from screen rendering (*ScR*). In the first case, the software is drawing the stimulus to display on a texture (via the GPU/CPU); in the second case, the pre-rendered texture is being displayed on the screen. Tagging rarely occurs after *ScR*, since this would require observing the colour of the pixels on the screen by means of a hardware or software component to ensure that the texture is displayed. More likely it would take place before or after *SoR*:

**Pipeline A**    *SoR → Tagging → ScR → Observation*

**Pipeline B**    *Tagging → SoR → ScR → Observation*

Once the stimulus is displayed on the screen, it can be observed by a photodiode. The response time of the photodiode is nearly instantaneous, less than one millisecond (ms). We define the latency ($L$) as the delay between the time the tagging command is sent and the detection of the stimulus by the photodiode.

B. Latency of tagging pipelines

In pipeline A, $L$ is mainly due to *ScR*, which is a linear function of the height $h$ of the stimulus on the screen, with the LCD screen refreshing from top to bottom. On a screen with a refresh rate (RR) of 60 Hz, *ScR* typically finishes within 20 ms:

*ScR*($h$) = $a * h + b$, where

$h$ is the position of the stimulus in screen height percentage, $a$ is the time taken to update the screen (16 ms by default), and $b$ is the time taken by a pixel to switch its colour (about 6 ms, Figure 1).

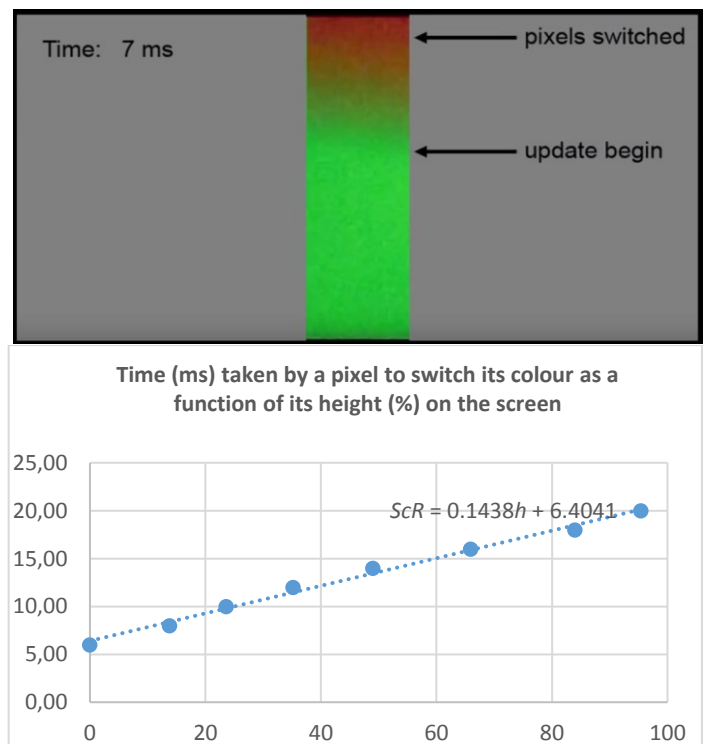

Figure 1. *LCD refreshes from top to bottom in linear time (14).*

*ScR* does not depend on *SoR*, but *SoR* does influence the *perceived ScR* (*PScR*). Ideally, when the number of frames per second (FPS) is equal to RR, a texture is displayed each time the screen is refreshed. However, when FPS is lower than RR, the time between two textures may be higher than 20 ms. In contrast, when FPS is higher than RR, this may cause *screen tearing*, that is, two frames rendering at the same time (15). Ideally, a stimulus should be drawn within a single frame and then displayed during *ScR*. This is generally the case when just one camera is rendering a virtual scene. However, it is not always so when the stimulus is rendered by more than one camera, as is the case in virtual reality (VR).

In pipeline B, the latency is due to both *SoR* and *PScR*.

**L(h) of Pipeline A**         *PScR*(h) + e

**L(h) of Pipeline B**         *SoR* + *PScR*(h) + e

In both conditions, *e* stands for hardly measurable delays, due to software and screen drivers, between the moment when the texture is ready and the moment when it is displayed. When estimating the latency with a photodiode, *e* also includes delays due to the sampling rate of the acquisition unit, the threshold used to detect the peaks of the photodiode and tagging signals, and the method used to remove the drift of the photodiode signal. We assume *e* and *SoR* to finish in near-constant time.

### C. Matrix of stimuli

When multiple stimuli are displayed, they are regrouped in a matrix (e.g. (16)). Thus, we can refer to a stimulus *S* using its position (*i*, *j*) in the matrix *I* x *J* rather than its height *h* on the screen. Therefore, we need to express *PScR* as a function of (*i*, *j*). The size of the screen is defined by *W* x *H*, corresponding to the actual width and height of the screen in pixels. *PScR* (*i*, *j*) may be seen as a compound of three functions:

1. *P*: *i*, *j* → *x*, *y* is the position of the stimulus in screen percentage, where (Figure 2)

$$x = \frac{(m_j + ju_j)}{W} \text{ and } y = \frac{(m_i + iu_i)}{H}, 0 < i < I \text{ and } 0 < j < J\ ;$$

2. *H*: *x*, *y* → *h* is the height of the stimulus on the screen, where *h* = *x* if the screen is turned 90° (as is the case for smartphone-based VR) and *h* = *y* otherwise; and

3. *PScR*(*h*) already defined

such that

*PScR*(*i*, *j*) = *PScR*(*h*) o *H*(*x*, *y*) o *P*(*i*, *j*)

with the following property (P), which corresponds to the delay between the observation of two stimuli

$$(P)\ |L(i_1, j_1) - L(i_0, j_0)| = |PScR(i_1, j_1) - PScR(i_0, j_0)|.$$

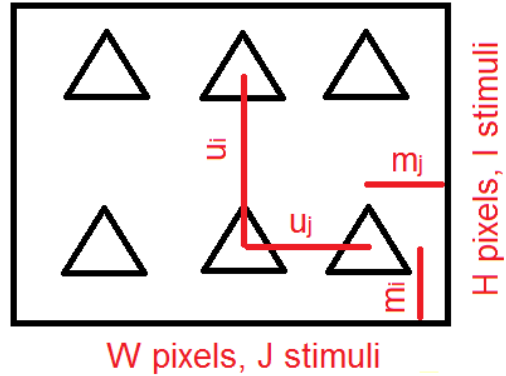

Figure 2. *Example of a matrix of stimuli with J stimuli in a row and I stimuli in a column. Ui is the distance between two adjacent stimuli in the same column. Uj is the distance between two adjacent stimuli in the same row. Mi and Mj are, respectively, the top or bottom and left or right margins.*

## III. Multiple display of a stimulus

As explained in section II.B, the *PScR* is probably modified when a stimulus is rendered by multiple cameras. This is always the case when using VR, for example, since it requires the display of two images for stereoscopic vision (Figure 3).

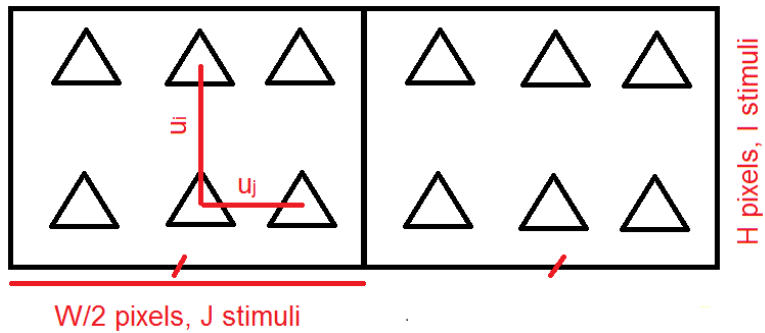

Figure 3. *Example of matrix of stimuli in VR having J stimuli in a line and I stimuli in a column.*

To avoid having a single stimulus being perceived as multiple stimuli, the delay between multiple appearances of the same stimulus on screen has to be as short as possible; that is, ideally *PScR ~ ScR*.

In practice, this is only the case if single-pass rendering is implemented. For instance, frameworks such as Unity (Unity, San Francisco, US), starting from version 2017.3, help to render the content of two stereoscopic cameras within a single frame (17).

Whenever this feature is not available, a solution is to take into account the *ScR* of just one camera, a camera being *a priori* rendered within a frame. However, it is not clear which camera has to be considered. At this point, we propose the *latency-of-first-appearance principle* (LOFAP); that is, the first appearance of the stimulus on screen induces the ERP, and thus the latency of this first appearance should be used to correct the offset of the ERP.

Under single-pass rendering or the LOFAP, the property (P) is simplified as follows:

(P)  $|L(i_1,j_1) - L(i_0,j_0)| = |ScR(i_1,j_1) - ScR(i_0,j_0)| = a\frac{u_i}{H}|i_1 - i_0|$ or $a\frac{u_j}{W}|j_1 - j_0|$ by symmetry.

Notice that at this point the latency between two stimuli depends only on the orientation of the screen (turned or not).

Nevertheless, the LOFAP does not correct for differences in amplitude which may result from multiple-pass rendering. Indeed, a high latency between the rendered content of the cameras may cause the stimulus to stay longer on the screen. In the case of visual stimulation such as the P300, however, this effect is minimal (18).

## IV. Experiment

We conducted an experiment to compare the ERPs produced by a user under two conditions: when the stimulus was displayed within a traditional PC application using a flat-panel display refreshing from top to bottom (condition 1), and when it was displayed on a smartphone-based VR immersion screen refreshing from left to right (condition 2). We used pipelines A and B for conditions 1 and 2, respectively. Pipeline A introduced less variability since it depended only on *ScR*. Unexpectedly, pipeline B gave better results (in terms of *jitter*) in condition 2. This can be attributed to the tagging method, which was partially asynchronous (Figure 4). Clearly this explanation has to be investigated further and the relationship between tagging pipelines and asynchronous tagging clarified.

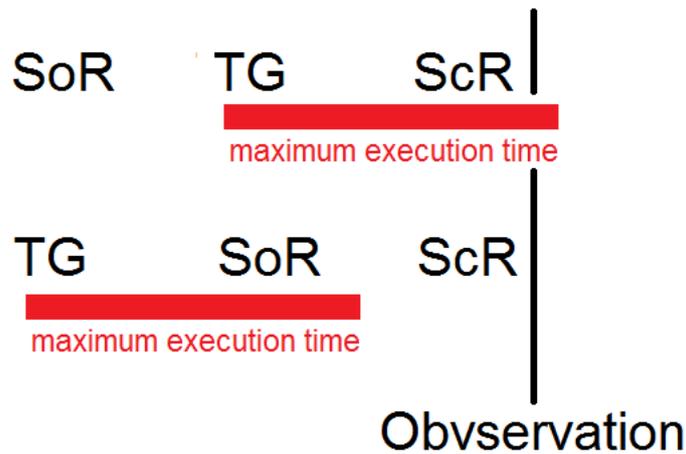

Figure 4. *Model for explaining the impact of asynchronous tagging: By moving asynchronous tagging before SoR, it finishes before ScR. Thus, the stimulus is observable subsequent to the tagging.*

We measured the latency between the tag and the appearance of a *centred* stimulus on the screen for each experimental condition. In condition 1, the latency was around 38 ms (standard deviation, or SD, = 5.3 ms). In condition 2, the stimulus appeared at two different locations on the screen, with latencies of 117 ms (SD = 5.8 ms) and 143 ms (SD = 6.6 ms) for the left and right parts of the smartphone screen, respectively. It was clear that the latency was higher in condition 1 than in condition 2. This was partially explained by the computation time due to the *SoR*, which was not present in pipeline A. Moreover, in VR two images are rendered at the same time, resulting in an extended computation time for the *SoR* in comparison to a traditional application. This impacted the *PScR*, as it became greater than 20 ms (143 ms – 117 ms > 20 ms), thus we applied the LOFAP and retained only the shorter latency (117 ms).

Having different latencies is a problem when comparing averaged ERPs in two conditions, since there is no way to know whether the difference in latency is physiological or is due to the tagging process. A solution is to subtract the estimated latencies we found in the two conditions from their respective ERPs. This removes the latency due to *SoR*, but not the latency due to *PScR*. In fact, the latency due to *PScR* in the averaged ERP depends on the location of the stimuli on the screen. If the distribution of the stimuli is *uniform*, their barycentre should be at the same location as the centred stimulus we used for estimating the latency. In practice, a centred stimulus in the matrix may differ from the barycentre if the matrix is even and/or if the number of stimuli is low (Figure 5).

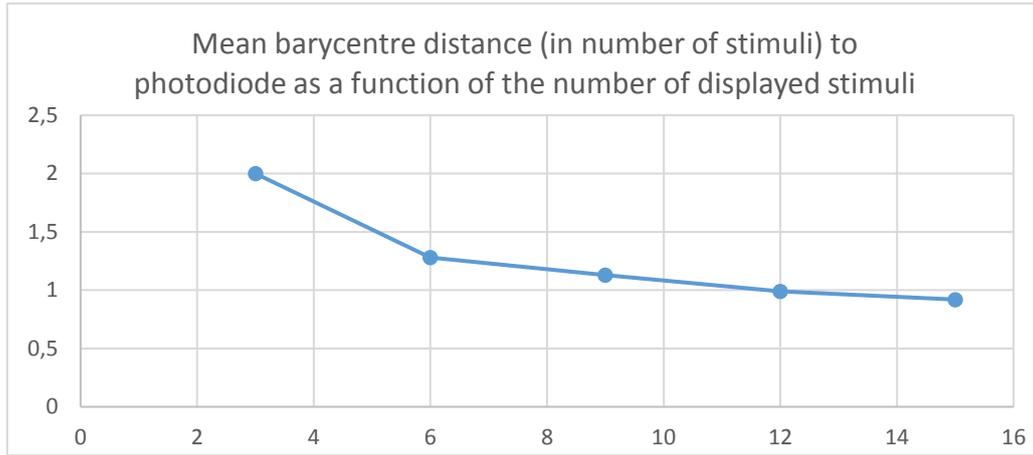

Figure 5. *Distance between the barycentre of the stimuli and the location of the photodiode. In this example, the photodiode was placed under the stimulus at position (2, 2) in a 6 x 6 matrix (indices start from zero). The* Y *axis represents the number of stimuli between the barycentre and the photodiode. The* X *axis represents the number of stimuli displayed during the experiment.*

In such case, the latency between the barycentre $(\bar{i}, \bar{j})$ and the centred stimulus $(i_0, j_0)$ can be estimated with the following:

(P) $|L(\bar{i}, \bar{j}) - L(i_0, j_0)| = a' \frac{u_j'}{W} |j_0 - \bar{j}|$ (condition 1) and $a \frac{u_i}{H} |i_0 - \bar{i}|$ (condition 2).

Notice that the latency depends only on the orientation of the screen—top-to-bottom (flat-panel display) or left-to-right (smartphone).

The values of *u, H*, and *W* were computed by measuring on the screens, and we used $a = a' = 16$ ms. Twelve stimuli were used for this experiment, with mean ($N = 10{,}000$) row and column distance to photodiode of 0 ms (SD = 0.4) and 0.2 ms (SD = 0.4), respectively, leading to

(P) $|L(\bar{i}, \bar{j}) - L(i_0, j_0)| = 0.54(1.23)$ ms in VR and $0.00(0.87)$ ms in PC with $L(\bar{i}, \bar{j}) \geq L(i_0, j_0)$ in the two conditions.

The average variability under the two conditions was reduced since *we used the same stimulus for the photodiode*. A maximum difference in latency of 0.9 ms (0.54 ms + 1.23 ms – 0 ms – 0.87 ms) between the two conditions can be observed due to the variability in the location of the barycentre. If the *uniform hypothesis is fulfilled* (as it is in the case using 12 stimuli) but the location of the photodiode is not known, the maximum difference in latency is about 2.6 ms for the matrix we used.

# V. Conclusion

In summary, tagging latency depends on the location of all stimuli on screen. This latency causes variability in the position of ERPs and can introduce a shift when averaging ERPs because not all stimuli have the same position on the screen. This variability is high when the number of stimuli is low. For a 6 x 6 matrix and 12 stimuli, the variability is only about 1 ms. Additional variability is introduced by positioning the photodiode at different locations when estimating latency. These errors can be corrected for if the location is known. When either the barycentre (of all the stimuli) or the photodiode location is not known, the maximum latency between two pixels on the screen should be lower than the refresh time of the screen, which is around 20 ms when the screen RR is 60 Hz. Uncertainty being cumulative, the maximum uncertainty is then in the range 30–40 ms when comparing two averaged ERPs (2 x 20 ms for a screen with RR of 60 Hz).

The subjects themselves are a source of variability (19,20). In fact, (20) reported that aircraft pilots were able to perceived between 12 and 220 FPS. The variability of the subject's perception of FPS is *a priori* negligible if the number of subjects is high and the experiment is paired (i.e., the paired ERPs in the two conditions are produced by the same subjects).

In conclusion, we recommend the following guidelines:

1) Use pipeline A with synchronous tagging when possible. With asynchronous tagging, pipeline B may lead to better performance.
2) The higher the RR, the better: whenever possible use a modern gaming monitor with a refresh rate of 140 FPS.
3) Use the VSync or, preferably, the NVIDIA GSync method. VSync allows the graphic card to match its FPS to the RR of the screen. Ideally GSync also lets the monitor vary its RR according to the FPS of the graphic card.
4) When multiple cameras are rendering, try to enable single-pass rendering. This helps to draw all cameras within a single frame (17). Otherwise, use the latency of the first displayed camera.
5) Make sure stimuli are uniformly distributed on the screen or that the barycentre of the displayed stimuli can easily be predicted.
6) Use the same photodiode location in all experimental conditions, close to the barycentre of the displayed stimuli.
7) Each subject has to be assessed under both conditions. Employ a substantial number of subjects to prevent variability in the perception of FPS by the human eye (increasing the sample size lower the variance). Exclude from the study subjects having trained vision, such as pilots or hard-core gamers.

# References

1. Luck SJ. Event-related potentials. In: APA handbook of research methods in psychology, Vol 1: Foundations, planning, measures, and psychometrics. Washington, DC, US: American Psychological Association; 2012. p. 523–46.

2. Wolpaw J, Wolpaw EW. Brain-Computer Interfaces: Principles and Practice. Oxford University Press, USA; 2012. 420 p.

3. Vidal JJ. Real-time detection of brain events in EEG. Proc IEEE. 1977 May;65(5):633–41.

4. Congedo M, Korczowski L, Delorme A, Lopes Da Silva F. Spatio-temporal common pattern: A companion method for ERP analysis in the time domain. J Neurosci Methods. 2016;267:74–88.

5. Congedo M. EEG Source Analysis [Internet] [Habilitation à diriger des recherches]. Université de Grenoble; 2013 [cited 2017 Apr 27]. Available from: https://tel.archives-ouvertes.fr/tel-00880483

6. Farwell LA, Donchin E. Talking off the top of your head: toward a mental prosthesis utilizing event-related brain potentials. Electroencephalogr Clin Neurophysiol. 1988 Dec;70(6):510–23.

7. Birbaumer N, Cohen LG. Brain–computer interfaces: communication and restoration of movement in paralysis. J Physiol. 2007 Mar 15;579(3):621–36.

8. Sellers EW, Donchin E. A P300-based brain–computer interface: Initial tests by ALS patients. Clin Neurophysiol. 2006 Mar 1;117(3):538–48.

9. Zander TO, Kothe C. Towards passive brain–computer interfaces: applying brain–computer interface technology to human–machine systems in general. J Neural Eng. 2011;8(2):025005.

10. Cattan G, Mendoza C, Andreev A, Congedo M. Recommendations for Integrating a P300-Based Brain Computer Interface in Virtual Reality Environments for Gaming. Computers. 2018 May 28;7(2):34.

11. Amin HU, Malik AS, Mumtaz W, Badruddin N, Kamel N. Evaluation of passive polarized stereoscopic 3D display for visual amp;amp; mental fatigues. In: 2015 37th Annual International Conference of the IEEE Engineering in Medicine and Biology Society (EMBC). 2015. p. 7590–3.

12. Pegna AJ, Darque A, Roberts MV, Leek EC. Effects of stereoscopic disparity on early ERP components during classification of three-dimensional objects. Q J Exp Psychol 2006. 2018 Jun;71(6):1419–30.

13. Käthner I, Kübler A, Halder S. Rapid P300 brain-computer interface communication with a head-mounted display. Front Neurosci. 2015;9:207.

14. renderingpipeline. LCD Screen Refresh in 1000 FPS [Internet]. [cited 2018 Nov 28]. Available from: https://www.youtube.com/watch?v=wts8f1bNnbo

15. Fedorov N. Frame Rate (FPS) vs Refresh Rate (Hz) [Internet]. AVADirect. 2015 [cited 2018 Nov 28]. Available from: https://www.avadirect.com/blog/frame-rate-fps-vs-hz-refresh-rate/

16. Guger C, Daban S, Sellers E, Holzner C, Krausz G, Carabalona R, et al. How many people are able to control a P300-based brain-computer interface (BCI)? Neurosci Lett. 2009 Oct 2;462(1):94–8.

17. Technologies U. Unity - Manual: Single Pass Stereo rendering (Double-Wide rendering) [Internet]. [cited 2018 Nov 28]. Available from: https://docs.unity3d.com/Manual/SinglePassStereoRendering.html

18. Lu J, Speier W, Hu X, Pouratian N. The Effects of Stimulus Timing Features on P300 Speller Performance. Clin Neurophysiol Off J Int Fed Clin Neurophysiol. 2013 Feb;124(2):306–14.

19. Humes LE, Busey TA, Craig JC, Kewley-Port D. The effects of age on sensory thresholds and temporal gap detection in hearing, vision, and touch. Atten Percept Psychophys. 2009 May;71(4):860–71.

20. Hagström R. Frames That Matter : The Importance of Frames per Second in Games [Internet]. 2015 [cited 2018 Dec 6]. Available from: http://urn.kb.se/resolve?urn=urn:nbn:se:uu:diva-263379